# Photoluminescence Spectra of Quantum Dots: Enhanced Efficiency of the Electron-Phonon Interaction

J. T. Devreese [1] (a), V. M. Fomin (a,b), V. N. Gladilin (a,b), and S. N. Klimin (a,b)

*(a) Theoretische Fysica van de Vaste Stof (TFVS), Universiteit Antwerpen (U.I.A.), B-2610 Antwerpen, Belgium*

*(b) Fizica Structurilor Multistratificate, Universitatea de Stat din Moldova, MD-2009 Chisinau, Moldova*

**Abstract.** A theory of photoluminescence in semiconductor quantum dots is developed which relies on two key ingredients. First, it takes into account non-adiabaticity of the exciton-phonon system. Second, it includes the multimode dielectric model of LO-phonons and of the electron-phonon interaction in confined systems. The role of non-adiabaticity is shown to be of paramount importance in spherical quantum dots, where the lowest one-exciton state can be degenerate or quasi-degenerate. For various quantum dot structures, the calculated intensities of phonon satellites as a function of temperature, excitation energy and size of quantum dots are in a fair agreement with experimental data on photoluminescence.

The interest in the properties of quantum dots is continuously growing because of the great prospects of these structures for optoelectronic applications. Numerous experiments, e. g., recent measurements [1 - 4] of the optical (photoluminescence, magneto-photoluminescence, resonant photoluminescence and photoluminescence excitation) spectra of self-assembled InAs/GaAs quantum dots reveal a remarkably high probability of phonon-assisted transitions. For some cases, attempts to interpret these experiments on the basis of the adiabatic theory meet considerable difficulties. For example, in spherical CdSe quantum dots, the calculated values of the Huang-Rhys parameter $S$ appear to be significantly (by one or two orders of magnitude) smaller than those derived from experiment. In order to achieve agreement with the experimental data, *ad hoc* mechanisms like charge separation by defect or surface states were introduced as a possible extrinsic origin of an increased Huang-Rhys parameter (see, e. g., [5]). However, some of the experimental spectra cannot be fitted by a Franck-Condon progression, predicted by the adiabatic theory, for any values of $S$, as seen from Fig. 1.

[1] Presenting author. Fax: +32-3-8202245; phone: +32-3-8202459; e-mail: devreese@uia.ua.ac.be

We have analyzed the photoluminescence spectra observed in different quantum-dot structures: ensembles of spherical CdSe quantum dots, self-assembled InAs/GaAs and CdSe/ZnSe quantum dots, brick-shaped InAs/GaAs quantum dots [6] created by AFM-mediated local anodic oxidation.

We show that the effects of non-adiabaticity provide an *intrinsic* reason to explain the observed surprisingly high intensities of the phonon satellites as well as the aforementioned discrepancies between experimental multiphonon photoluminescence spectra and those calculated within the adiabatic approach. The non-adiabaticity of the exciton-phonon system in quantum dots opens different additional channels of the phonon-assisted optical transitions [7]. Taking into account those channels allows us to adequately treat experimental data on photoluminescence of quantum dots. This is illustrated by the spectra calculated in Ref. [7] and shown in Fig. 1 (for colloidal spherical CdSe quantum dots) and in Fig. 2 (for spherical CdSe quantum dots embedded into borosilicate glass).

In order to treat the fine structure of the phonon satellites, we have obtained the optical-phonon modes and the Hamiltonian of the electron-phonon interaction characteristic for quantum-dot structures with various geometric and material parameters. These calculations are based on the multimode dielectric model [10, 11], which is another cornerstone of our theory. Our model explicitly takes into account the fact that the number of vibrational modes in a nanostructure is finite. Representing the relative ionic displacement as a finite sum over basis vectors, which satisfy specific mechanical boundary conditions, we solve jointly the generalized Born-Huang equation [10, 12] and the Maxwell equations with the electrostatic boundary conditions. It appears that, in the case of dispersive phonons, the vibrational modes in a quantum dot cannot be classified into bulk-like or interface modes, but have a mixed character. In terms of resonant relaxation of excitons with participation of these hybrid phonons, we interpret in Ref. [11] the fine structure of phonon replicas, which was recently observed in photoluminescence spectra of self-assembled CdSe/ZnSe quantum dots [13, 14].

In Fig. 3, the theoretical results of Ref. [11] for the phonon-assisted photoluminescence spectra are shown for two ensembles of cylindrical CdSe quantum dots with fixed height $h$ and with various diameters $d$. As follows from this figure, three peaks, which belong to the first phonon replica in the measured photoluminescence spectra of CdSe quantum dots [13, 14], can be attributed to spatially extended ZnSe-like phonon modes, localized CdSe-like modes and spatially extended CdSe-like modes (in order of increasing detection energy $\hbar\Omega$). Calculated positions and relative heights of peaks induced by spatially extended ZnSe- and CdSe-like phonon modes are in a fair agreement with the experimental data [14]. The experimental peaks, which can be attributed to localized phonons, are characterized by a rather large frequency spread when compared with theoretically calculated peaks. This may be related to an inhomogeneous strain distribution as well as to fluctuations of the chemical composition in the experimental samples.

The authors acknowledge fruitful collaboration with E. P. Pokatilov and S. N. Balaban. We thank F. Henneberger and M. Lowisch for providing us with their experimental data and for useful discussions. This work has been supported by the I.U.A.P., GOA BOF UA 2000, F.W.O.-V. projects Nos. G.0287.95, 9.0193.97 and the W.O.G. WO.025.99N (Belgium).

**Figure captions**

Fig. 1. Photoluminescence intensity of spherical CdSe quantum dots with the average radius $\langle R \rangle = 1.2$ nm as a function of the difference between the emission frequency $\Omega$ and the excitation frequency $\Omega_{\mathrm{exc}}$. The dashed curve represents the experimental data [8], the dash-dotted curve displays a Franck-Condon progression with the Huang-Rhys parameter $S = 0.06$ calculated in [5], the dotted curve shows another Franck-Condon progression with $S = 1.7$, which is obtained by fitting the ratio of intensities of two-phonon and one-phonon peaks to the experimental value, and the solid curve results from the present theory. (From Ref. [7].)

Fig. 2. Photoluminescence spectra of CdSe quantum dots at different average radii and excitation energies. Solid lines represent the families of experimental time-resolved photoluminescence spectra [9] measured at different time intervals between the pumping pulse and the measurement, the upper curve corresponding to the time interval equal to 0. Theoretical results [7] are displayed for the equilibrium-photoluminescence spectra (dashed curves) and for the case of slow relaxation of the exciton energy (dotted curves).

Fig. 3. Calculated photoluminescence spectra [11] of CdSe quantum dots with the height $h = 2$ nm (the solid line) and $h = 1.8$ nm (the dotted line) on a 2ML remnant layer compared to the experimental data for a CdSe quantum dot structure formed by thermally activated surface reorganization of an initially uniform 3 ML CdSe film (after Lowisch *et al.* [14]). The dashed and dot-dashed lines correspond to the excitation energy $\hbar\Omega_{\mathrm{exc}} = 2.495$ eV and $\hbar\Omega_{\mathrm{exc}} = 2.465$ eV, respectively. For clarity, the experimental spectra are vertically shifted.

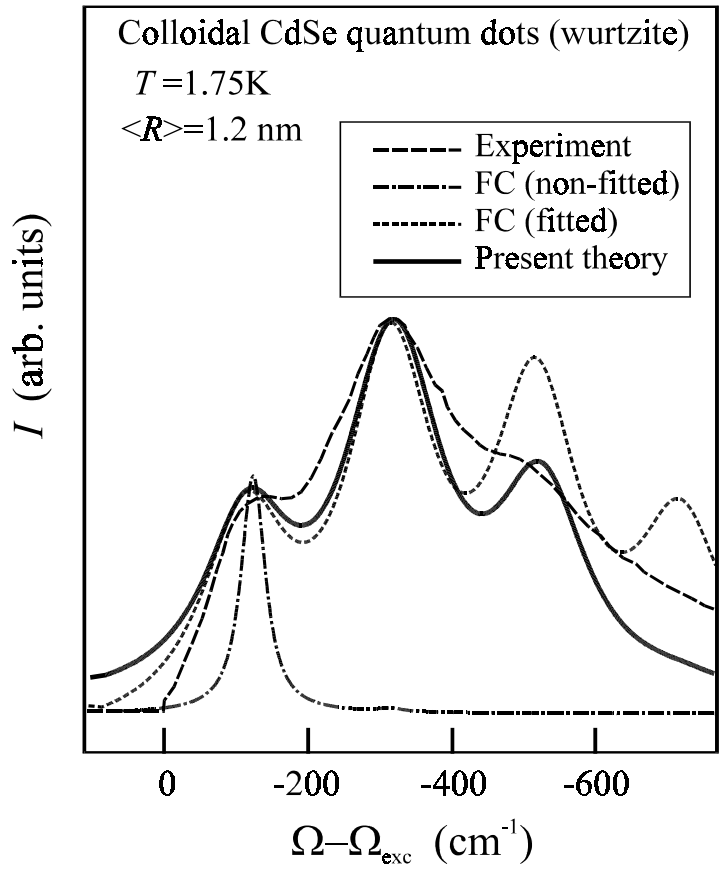

Fig. 1     J. T. Devreese, V. M. Fomin, V. N. Gladilin, and S. N. Klimin

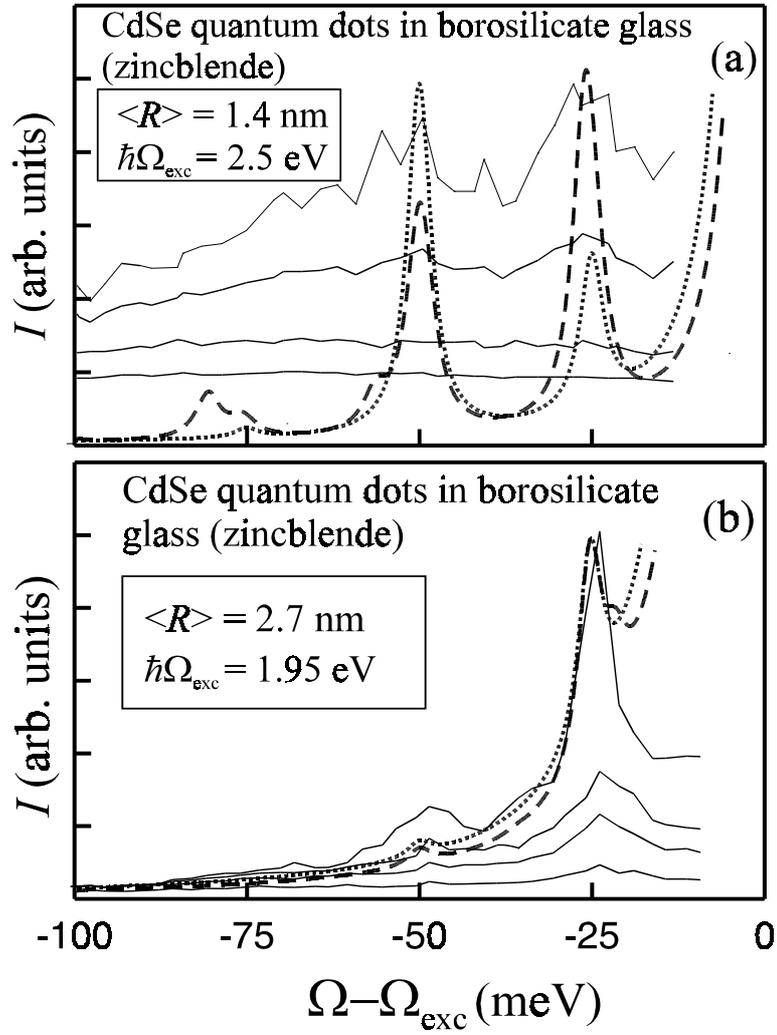

Fig. 2          J. T. Devreese, V. M. Fomin, V. N. Gladilin, and S. N. Klimin

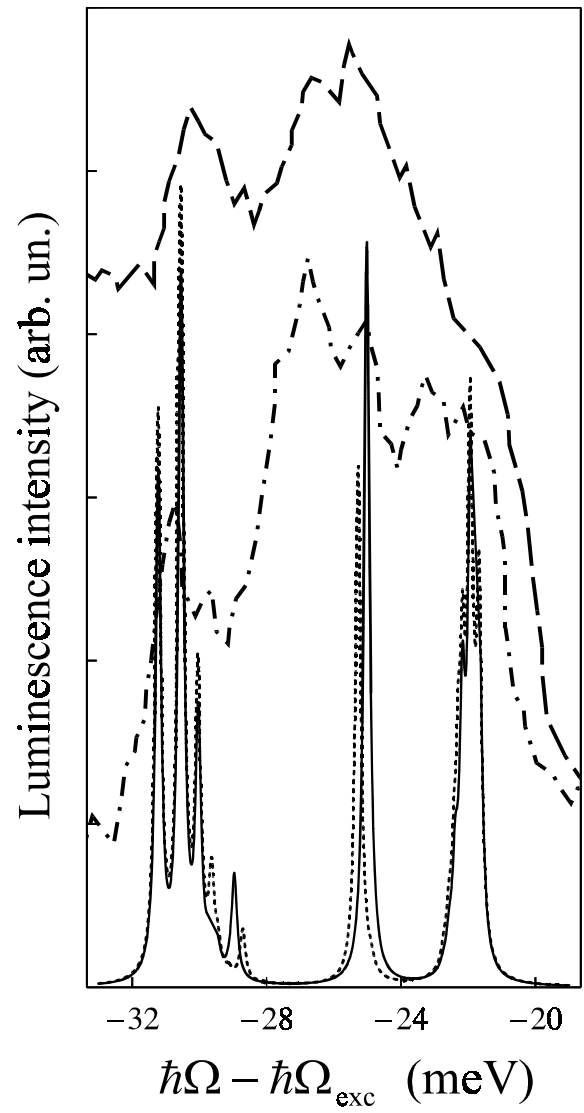

Fig. 3            J. T. Devreese, V. M. Fomin, V. N. Gladilin, and S. N. Klimin